# Structural, dielectric and energy storage properties of BaO–Na$_2$O–Nb$_2$O$_5$–P$_2$O$_5$ glass-ceramics

*Abderrahim Ihyadn,*[a1] *Abdelilah Lahmar,*[b] *Igor Luk'yanchuk,*[b,c] *Daoud Mezzane,*[a,b] *Lahcen Bih,*[d,e] *Abdelhadi Alimoussa,*[a] *M'barek Amjoud*[a] *& Mimoun El Marssi*[b]

[a] *IMED-Lab, Cadi Ayyad University, Marrakesh, 40000, Morocco*
[b] *LPMC, University of Picardy Jules Verne, Amiens, 80039, France*
[c] *Physics Faculty, Southern Federal University, Rostov-on-Don, 344090, Russia*
[d] *Département Matériaux et Procédés, ENSAM Meknès, Université Moulay Ismail, Meknès, Maroc*
[e] *Equipe Physico-Chimie la Matière Condensée (PCMC), Faculté des Sciences de Meknès, Université Moulay Ismail, Maroc*

*A series of (1−x)[(2BaO–0·5Na$_2$O)–1P$_2$O$_5$] –xNb$_2$O$_5$ (BNPN, x=0·41, 0·43, 0·45, 0·48) glass-ceramics based on phosphate glasses have been prepared via a controlled-crystallisation route. The structure, dielectric properties, interfacial polarisation and energy storage properties were systematically investigated. The x-ray diffraction results showed the simultaneous presence of Ba$_2$NaNb$_5$O$_{15}$ tungsten bronze structure (TTB) and the NaNbO$_3$ perovskite. A stable dielectric constant over a temperature range from 25–200°C, low dielectric losses less than 0·03 and excellent frequency stability at room temperature were obtained. The decrease of niobium content promoted the TTB crystallisation with improvement of the high dielectric properties of the system. The optimum of the dielectric constant and recoverable energy storage density were obtained for BNP41 crystallised at 1000°C. Analyses of the complex impedance indicated that the niobium content and crystallisation temperature affect the interfacial polarisation.*

## 1. Introduction

In recent years, energy materials devices such as capacitor and electrochemical cell have attracted tremendous attention. Among various capacitor materials, dielectric glass ceramics are presently the materials of choice for pulse power technology applications and have received much attention.[1] Although the traditional ferroelectric ceramics have been used in capacitors because of their high dielectric constant, the low breakdown strength (BDS) and high remnant polarization limits their applications.[2] To overcome the problem, the ferroelectric glass-ceramics were suggested as alternative. Indeed, ferroelectric glass-ceramics can be defined as a composite of a glass phase with high breakdown strength and crystalline phase with high dielectric constant.[3] In recent years, niobate glass-ceramics have been proposed as a good alternative for high energy storage application as dielectric capacitor.[4]

In fact, high dielectric constant, high breakdown strength, and low dielectric loss are needed to achieve a high energy density.[5] It should be noted that the synthesis of high permittivity glass-ceramics has been the subject of many studies.[4,6] Glass materials with high dielectric constant can be made either by adapting the chemical composition[7] or by the precipitation of crystalline phase with a high dielectric constant.[6] On the other hand, numerous studies have found that the control of the crystallisation temperature dramatically improves the dielectric properties of the glass-ceramics.[4]

In this context, niobium (Nb)-related dielectrics recently have absorbed much attention because of their importance in development of lead (Pb)-free glassy ferroelectrics. Many ferroelectric glass-ceramics use niobium oxide as a key constituent in the SiO$_2$ and/or B$_2$O$_3$ precursor glass, because it has clearly shown its primordial role in obtaining good ferroelectricity in niobite materials. Different types of ferroelectric glass-ceramic systems such as (BaO, Na$_2$O)–Nb$_2$O$_5$,[8] (SrO, BaO)–Nb$_2$O$_5$,[9] (Na$_2$O, SrO)–Nb$_2$O$_5$,[10] (K$_2$O, SrO, BaO)–Nb$_2$O$_5$[11] have been investigated widely. For instance, Takashi *et al*[12] reported that (BaO, Na$_2$O)–Nb$_2$O$_5$–P$_2$O$_5$ glass-ceramics crystallised in phosphoniobate system. They found that both dielectric constant and refractive index increased by the nanocrystallisation. Bih *et al*[13] investigated *y*(Ba$_{2·15−x}$Na$_{0·7+x}$Nb$_{5−x}$W$_x$O$_{15}$)–(1−*y*)P$_2$O$_5$ glass-ceramics, noticed the formation of glass-ceramics with TTB

A. IHYADN ET AL: STRUCTURAL, DIELECTRIC AND ENERGY STORAGE PROPERTIES OF $BaO-Na_2O-Nb_2O_5-P_2O_5$phases and the behaviour of high dielectric constant was attributed to highly polisable Ba, Nb and W ions.

Furthermore, in terms of energy storage application, $(BaO, Na_2O)-Nb_2O_5$ system deserves a special interest, because such a system might contain $Ba_2NaNb_5O_{15}$ (BNN) with TTB-structure that combines a high dielectric constant (~240) and high Curie temperature of about 600°C (large ferroelectric domain comparing to the $BaTiO_3$). Nevertheless, only few investigations of phosphate glass-ceramics are reported dealing with this topic.

Recall that BNN is considered as only one point of a solid solution, isolated within the ternary diagram $BaO-Na_2O-Nb_2O_5$, with tetragonal tungsten bronze structure type. Further, the variation in stoichiometry near BNN is reported either by Jamieson et al. $(Ba_{4+x}Na_{2-2x}Nb_{10}O_{30})$ or by Singh et al $(Ba_{4+x}Na_{2-2x}Nb_{10}O_{30})$.[14,15]

Therefore, the purpose of the present work is the stabilisation of this phase in a glassy matrix in order to increase the breakdown field and improving energy storage density. As matrix, we interest on lead-free $(1-x)[(2BaO-0.5Na_2O)-1P_2O_5]-xNb_2O_5$ (BNPN, $x$=0·41, 0·43, 0·45, 0·48) phosphoniobate glass-ceramics prepared via body crystallisation method. The emphasis is placed on the effect of the $Nb_2O_5$ content on thermal, structural and electric properties of BNPN glass-ceramics. The crystallisation temperature effect on dielectrics and energy storage properties was investigated.

## 2. Experimental procedures

Samples of the BNPN glass–ceramics with the nominal composition $(1-x)[(2BaO-0.5Na_2O)-1P_2O_5]-xNb_2O_5$ (BNPN, $x$=0·41, 0·43, 0·45, 0·48), designated as BNP41, BNP43, BNP45, BNP48, respectively, were prepared by the controlled crystallisation method. The powders containing $Na_2CO_3$, $BaCO_3$, $Nb_2O_5$ and $(NH_4)H_2PO_4$ were ball mixed for 4 h in a high density polyethylene bottle with ethanol as milling media for homogenous mixing, and then were dried at 100°C for 24 h. The dried powders were put into aluminia crucible and melted at 1350°C for 40 min in air. The melt was quickly poured into a preheated steel plate (200°C) and pressed with another steel plate. The tempered glass was cut into sheet-shaped samples. The casting glass was then annealed for 1 h at $T_g$–50 to relax the internal thermal stresses. Thereafter, all glass samples were crystallised at 760°C for 10 h in air with the heating rate of 10°C/min. Furthermore, the BNP41 glass was crystallised at 800, 900 and 1000°C for 10 h.

The crystallisation behaviour of the glass was analysed by differential scanning calorimetry (DSC, Model SDT-Q600) with a rate of 10°C/min$^{-1}$ using glass powders. The x-ray diffraction (XRD, Panalytical™ X-Pert Pro spectrometer) having a Cu K$_\alpha$ radiation ($\lambda$=1·54182 Å) were used to confirm the amorphous nature of the prepared glasses and to investigate the phase structure of glass-ceramics powders. The density ($d$) of the monolithic glass-ceramics was determined at room temperature by Densimeter Model H-300™ S, using the Archimede's method with distilled water as an immersion fluid. At least three samples of each glass-ceramic composition were selected to determine the density. The Raman spectroscopy was carried out on monolithic samples using (confotec MR520™) with a green excitation laser of 532 nm and a high speed digital colour CCD camera. Infrared spectra of the samples were recorded on a Bruker Vertex 70 Fourier transform infrared spectroscopy (FTIR) spectrometer. The measurements were made on a glass powder dispersed in 1 wt% KBr pellets. The electrical and dielectric measurements were carried out on silver paint coated samples using an impedance analyser (LCR meter hp 4284A 20 Hz–1 MHz). Complex impedance spectra were carried out a wide temperature range of 400–480°C and in the frequency range of 20 Hz to 1 MHz. This impedance response obtained could be ideally simulated by using a simple parallel resistance–capacitance (R–C) circuit. The contributions of the resistive ($R$) and capacitive ($C$) elements will indicate the appropriate regions of the different semicircular arcs. For better data fitting, the constant phase elements (CPE) instead of capacitances ($C$) were used in the equivalent circuit, due to the non-ideal Debye relaxation model.[16] A ZVIEW® Software was used to fit the experimental data obtained. Finally, the energy storage behaviour was investigated on silver paint coated samples using $P–E$ hysteresis loops, which were measured by a ferroelectric tester (the TF Analyzer 3000) at room temperature and at the frequency of 1 kHz.

## 3. Results and Discussion

### 3.1. Differential scanning calorimetry (DSC)

The DSC curves of as-quenched BNPN glasses at the heating rate of 10°C/min are illustrated in Figure 1. $T_g$ corresponds to the glass transition temperature. It is observed that the glass transition temperature ($T_g$) increased with the increase of $Nb_2O_5$ content. This result could be ascribed in the fact that to the niobium addition to the glasses. So, the (P–O–P) bonding will be broken and transferred to asymmetric bridging oxygen (Nb–O–P) and (Nb–O–Nb) bonding. This leads to the network crosslinking density, since the strength of these bonding are stronger than that of (P–O–P) bonding.[17] The presence of these vibrational modes is confirmed by IR and Raman as we will see in the next section. In addition, all DSC curves show two exothermic peaks, one $T_{p1}$ in the range of 710–755°C and the other $T_{p2}$ in the range of 778–803 °C. $T_{p1}$ and $T_{p2}$ correspond to the precipitation temperatures of ceramic phases.[18] Temperatures characteristics ($T_g$, $T_{p1}$, and $T_{p2}$) of the glasses as well as densities of glass-ceramics are collected in Table 1. The presence of such two anomalies crystallisation peaks $T_{p1}$ and $T_{p2}$ is an indication of the formation of two crystalline phases from the glass matrix. According to the DSC results, the glasses were crystallised at the 760°C for 10 h in order to obtain glass-ceramics.

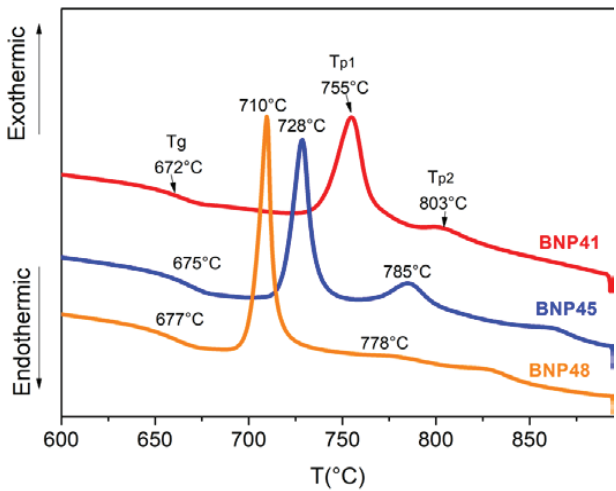

Figure 1. Differential scanning calorimetry (DSC) curves of the BNPN glasses powders with a heating rate of 10°C/min [Colour available online]

The values of the density of BNPN glass-ceramics are presented in Table 1. One can observe that the density increases with increasing Nb$_2$O$_5$ content. The density is increased from 3·65 to 3·98 g/cm$^3$. This result indicates that the glass structure becomes more tightly packed with niobium content. Although the contribution of the higher molecular weight of Nb$_2$O$_5$ cannot be neglected. This means that the $T_g$ increase with niobium content as observed in DSC results.

### 3.2. Room temperature x-ray diffraction analysis (XRD)

Figure 2 shows XRD patterns of the glass-ceramics samples crystallised at 760°C for 10 h. By the comparison to ICDD card, XRD spectra reveal that two main phases are formed: Ba$_2$NaNb$_5$O$_{15}$ with tetragonal tungsten bronze structure (TTB) (00-013-0575) and NaNbO$_3$ with perovskite structure (01-074-2454).[19,20] These results are consistent with those previously reported by Takahashi.[12] Moreover, the details of the evolution of these phases have been reported elsewhere.[19] Besides, Ba$_2$NaNb$_5$O$_{15}$ (BNN) was observed as a major phase and NaNbO$_3$ (NN) as a minor phase will be obtained. It can be clearly found that the intensity of these phases increases continuously with decreasing of $x$. It seems that the decrease of $x$, results in the increase of the formation of crystalline phase, which may result in high dielectric constant of glass-ceramics.

Table 1. Results of DSC analyses of BNPN glasses and values of density of glass-ceramics BNPN

| x | Sample name | $T_g$ (°C) | $T_{p1}$ (°C) | $T_{p2}$ (°C) | Density (g/cm$^3$) |
|---|---|---|---|---|---|
| 0·41 | BNP41 | 672 | 755 | 803 | 3·65±0·05 |
| 0·45 | BNP45 | 675 | 728 | 785 | 3·86±0·06 |
| 0·48 | BNP48 | 677 | 710 | 778 | 3·98±0·04 |

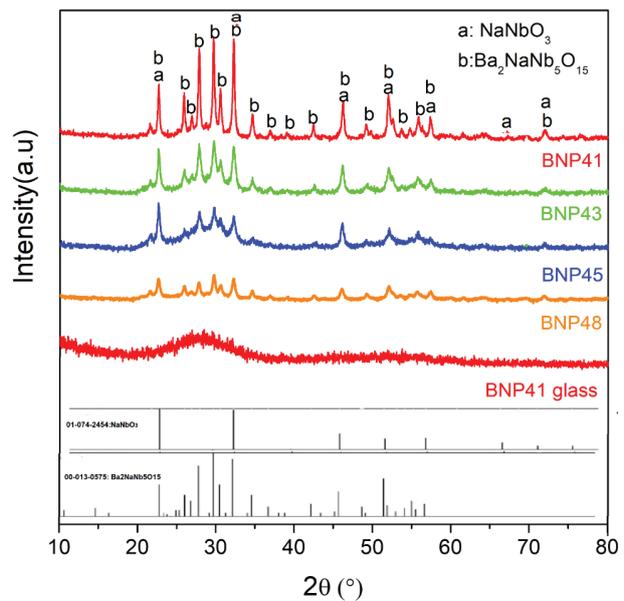

Figure 2. XRD patterns of BNPN glass-ceramics specimens and the BNP41 glass [Colour available online]

Besides, it is observed that, the XRD patterns of glass BNP41 (Figure 2) consist of broad halo peaks at low diffraction angles, which confirm the amorphous nature of the glass. Furthermore, all glasses show same results. It can be seen also that the BNP41 glass amorphous becomes a glass-ceramics after the heat treatment at 760°C for 10 h.

### 3.3. Infrared and Raman spectroscopy

The obtained FTIR spectra in the wave number range of 1300–300 cm$^{-1}$ are presented in Figure 3(a). The bands of FTIR spectra are not very well resolved because of the typical broadening observed. A deconvolution method with a Gaussian function was used to identify those peak. The attribution of the absorption bands is done by comparing our results with data provided in the literature.[21,22] The absorption band at 1268 cm$^{-1}$ correspond to symmetric stretching of P–O–P group. The weak absorption band observed in the range 1220–1195 cm$^{-1}$ is attributed to the asymmetric stretching vibrational of P=O bond, $\nu_{as}$(P–O); this band may also consist of bands due to asymmetrical vibrations of (PO$_2$)$^-$ mode, $\nu_{as}$(PO$_2$)$^-$, in metaphosphate Q$_2$ units. The absorbance band at around 1118 cm$^{-1}$ corresponds to the asymmetric stretching vibration of P–O bonds as well as the asymmetric stretching vibrations of (PO$_3$)$^{2-}$ of pyrophosphate groups (or Q$_1$ units).[23] The absorbance band observed at 1045 cm$^{-1}$ is assigned to orthophosphate PO$_4^{3-}$ groups (or Q$_0$ units). The absorption band at 910–960 cm$^{-1}$ is due to asymmetric stretching vibration of P–O–P bonding, denoted as $\nu_{as}$(POP).[24] Phosphate glasses can be produced with relatively high amounts of niobium by increasing the cooling rate without crystallisation. Niobium is a glass former, is located in octahedral sites and replaces tetrahedral phosphate groups by P, Nb, and O linked mixed chains. The density increases from 3·32 to 3·73g/cm$^3$, the elastic modulus increases from 56 to 78 GPa, and the linear thermal expansion coefficient decreases from 24·7×10$^{-6}$ to 7·9×10$^{-6}$ °C$^{-1}$ as the amount of Nb increases. The band at 724 cm$^{-1}$ assigned to symmetric stretching vibrations of P–O–P bridging oxygen atoms, denoted as $\nu_s$(POP). The bands at around 650 cm$^{-1}$ is assigned to Nb–O bonding from the NbO$_6$ units, and the intensity of this band increases with increasing Nb$_2$O$_5$ content. An absorption band around 820 cm$^{-1}$ corresponds to Nb–O bonding associated with NbO$_4$ units.

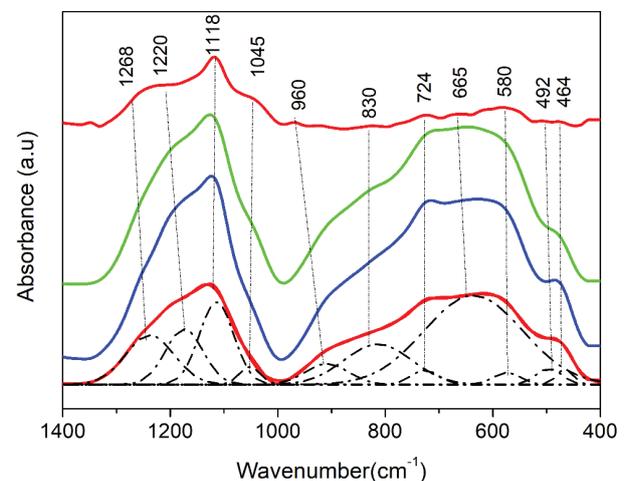

Figure 3. FTIR spectra of the studied BNPN glass-ceramics [Colour available online]

The decrease of the intensity of absorption around 1220 and 1268 cm$^{-1}$ with the increase of $x$ indicates the decrease of P–O–P and the P=O double bond in the BNPN glass-ceramics[24] thermal analysis, Raman and infrared spectroscopy. Phosphate glasses can be produced with a relatively high amount of niobium by increasing the cooling rate without crystallisation. Niobium is a glass former, is located in octahedral sites, and replaces tetrahedral phosphate groups by P, Nb and O linked mixed chains. The density increases from 3·32 to 3·73 g/cm$^3$, the elastic modulus increases from 56 to 78 GPa, and the linear thermal expansion coefficient decreases from 24·7×10$^{-6}$ to 7·9×10$^{-6}$ °C$^{-1}$ as the amount of Nb increases. The glass transition temperature ($T_g$, is weakened when Nb$_2$O$_5$ content increases from 41 to 48 mol%. In contrast to this band change, a new band gradually appears at about 465 cm$^{-1}$, which can be attributed to the bending vibration of the O–Nb and O–Nb–O–P links.[25] The absorptions associated with the nonbridging oxygen (NBO) (P–O$^-$) located at around 1118 cm$^{-1}$ also increases with the presence of the Nb$_2$O$_5$. The band centered at 492 cm$^{-1}$ whose intensity decrease with $x$, is related to a harmonic vibration assigned to bending of O–P–O and O=P–O bonds.[26]

From the FTIR spectroscopy, we infer that niobium is an intermediate glass in phosphate glasses since the addition of Nb leads to the formation of groups with O–Nb and O–P bounds. The results indicate that a more compact and cross-linked glass network is formed when P$_2$O$_5$ is increasingly replaced by Nb$_2$O$_5$ which consistent with DSC results. In fact, the increase in $T_g$ values is due to the enhancement of the bond strength and the reticulation of the glassy-matrix. The denser and more integrated glass network with the increase in Nb$_2$O$_5$ is also proved by the increase in glass-ceramics density.

Figure 4 shows the Raman spectra (100–1200 cm$^{-1}$) of the glass-ceramics, and the deconvoluted spectra of BNP48 glass-ceramic. The bands at around 202 and 243 cm$^{-1}$ correspond to $\delta$(O–P–O) of PO$_4$ units and/or $\delta$(O–Nb–O) deformation modes of NbO$_6$.[27,28] The strong band at 640 cm$^{-1}$ is attributed to $\nu_s$(Nb–O) vibrations in NbO$_6$ octahedrons.[29,30] The bands at around 730 cm$^{-1}$ are due to bridging oxygen P–O–P stretching modes. The P–O–P bonding strength decreases by increasing Nb$_2$O$_5$. The 889 cm$^{-1}$ band can be attributed to the distorted octahedron [NbO$_6$] having at least one short Nb–O bond pointing to a modifier ion.[27,29]

By increasing Nb$_2$O$_5$ content, the niobium ions tend to play an intermediate role by breaking P–O–P and O–P–O bonds forming [NbO$_6$] structure. These results are in ageement with those of DSC and FT-IR.

In the Raman spectrum of the BNN crystal, characteristic bands appear around 280 and 650 cm$^{-1}$,[31] which are in agreement with our results.

### 3.5. Dielectric properties

Figure 5 presents the temperature dependence of dielectric permittivity of BNPN glass-ceramics at 10 kHz. Two distinct behaviours of temperature dependence of dielectric permittivity are distinguished. The first one is in the range of 25–200°C and the second one in the range of 200–460°C. It can be seen at the first behaviour that, all the samples exhibited a stable dielectric constant over the temperature. Besides, the phosphate glass-ceramics displayed moderate values of the dielectric constant (50–80) at room temperature as compared to other silicate glass-ceramics.[19,20] It is observed also that the dielectric permittivity increase with the decrease of niobium content. This observed trend could be attributed to the formation of more compact and integrated glass structure leading to a decrease of the dipole polarisation.[26] This result is consistent with our results obtained in infrared and Raman spectroscopy. Furthermore, the enhanced relative permittivity is mainly due to the precipitation of the Ba$_2$NaNb$_5$O$_{15}$ and NaNbO$_3$ with high $\varepsilon_r$ in the glass matrix as observed in XRD results. Moreover, at room temperature the values of the dielectric loss were ranging ~0·004–0·03 and the lower dielectric loss 0·004 was obtained for BNP41.

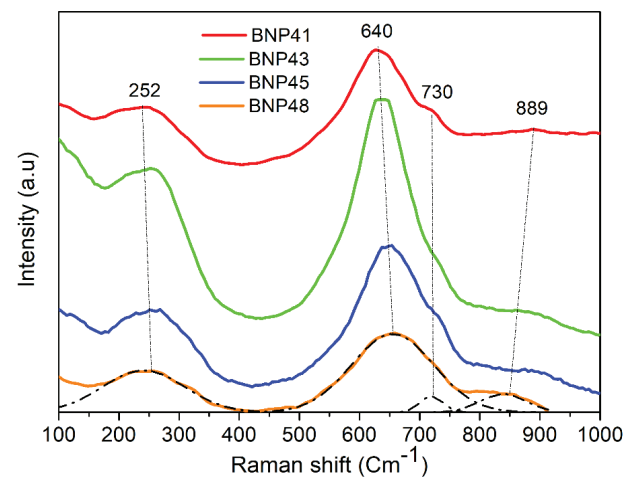

*Figure 4. Raman spectra of the studied BNPN glass-ceramics [Colour available online]*

On the other hand, at high temperature in the second behaviour, the dielectric constant tended to increase with the increase in $x$. This result can be interpreted as an increase of high polarisability of Nb$_2$O$_5$ with increase of temperature.

The dielectric properties of the BNPN glass-ceramics at room temperature as a function of frequency are presented in Figure 6. A good frequency stability of dielectric permittivity can be observed for all the samples. Furthermore, it is illustrated in Figure 6 that the dielectric permittivity of BNP41 glass increases from ~34 to ~75 for BNP41 glass-ceramics.

In order to obtain a dielectric with high permittivity, the effect of crystallisation temperature was discussed on BNP41 sample, where the optimal dielectric constant of ~75 and the lower dielectric losses of ~0·004 were obtained. Figure 7 shows the temperature dependence of both dielectric constant and dielectric loss performed for BNP41 glass crystallised at 760, 800, 900 and 1000°C. It can be seen that the dielectric constant increases with the increase of the crystallisation temperature. For example, the dielectric constant of glass BNP41 at 10 kHz increases from ~34 to 75 at 760°C and to 272 when the crystallised at 1000°C. This variation could be attributed to the increase of the amount of Ba$_2$NaNb$_5$O$_{15}$ and NaNbO$_3$ phases, known for high $\varepsilon_r$, while increasing crystallisation temperatures.[4] Henceforth, the dielectric loss exhibits a small increase trend with the crystallisation temperatures increasing, such as 0·004, 0·007, 0·022, and 0·011 for 760, 800, 900, and 1000°C at 10 kHz, respectively. This increase could be associated to the increase of the degree of the crystallinity. Albeit, the dielectric loss still below 0·02.

### 3.5. Impedance spectroscopy analysis

The complex impedance spectroscopy and the equivalent circuit of glass-ceramics content are shown in Figure 8. Generally, glass-ceramics can be considered as a system combining the crystalline and the glass phases. The crystallites and the glass can have a different structure from each other; also their properties can be different from that of the glass/crystal interface.[32,33] Thus, the contributions of the glassy matrix, crystal phase and the glass/crystal interface can be distinguished.

$Z'$–$Z''$ plot for BNPN glass-ceramics in Figure 8(a) shows a single semicircle at 480°C in the impedance spectra measured from 20 Hz to 1MHz. The resistance and capacitance values of equivalent circuits are tabulated in Table 2. It is noted from this Table that the estimated capacity for the first and third equivalent circuit is in the order of $10^{-10}$ and $10^{-11}$ F, which is typical of bulk electrical processes, whereas, the second circuit revealed capacitance values in the order of $10^{-9}$ F, which is characteristic of interfacial processes.[34] Moreover, it is found that the magnitude of resistance corresponding to $R_1$ is one or two order higher than other resistances ($R_2$ and $R_3$) for the all BNPN glass-ceramics. This means that the first arc corresponding to $R_1$-$CPE_1$ at high frequency region should represent the contribution of crystalline phases.[35] Hence, the first one $R_1$-$CPE_1$ circuit is attributed to the contribution of crystallites appearing at high frequencies followed by the second $R_2$-$CPE_2$ associated with the glass–crystal interface and the third $R_3$-$CPE_3$ ascribed to the glassy matrix at lower frequencies. It is observed that the values of $Z'$ are higher for BNP43 when compared with the other samples at similar temperatures. This is may be due to the blocking of mobile ions by the crystals in the network.[36]

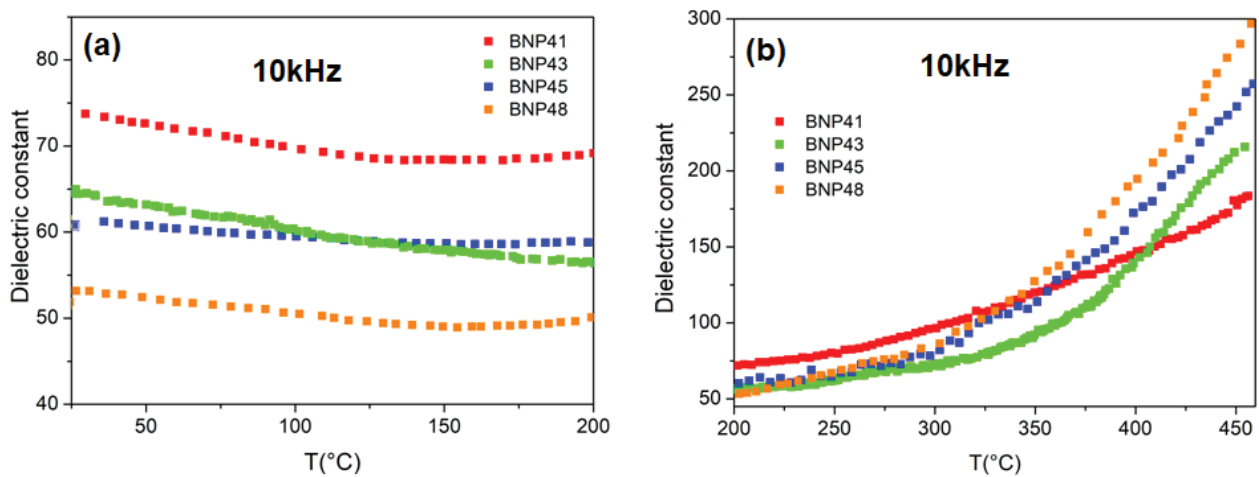

Figure 5. Temperature-dependence of dielectric permittivity of BNPN glass-ceramics at 10 kHz :(a) first range, (b) second range [Colour available online]

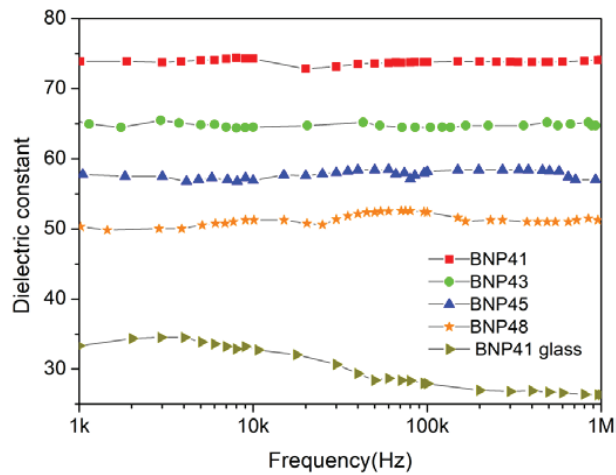

Figure 6. Frequency dependence of dielectric permittivity of BNPN glass-ceramic and BNP41 glass at room temperature [Colour available online]

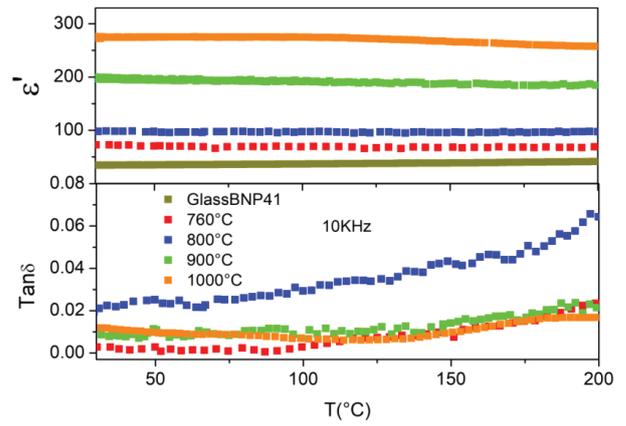

Figure 7. Temperature dependence of dielectric permittivity and dielectric loss of BNNP41 glass-ceramics at 10 kHz [Colour available online]

Figure 8(b) shows impedance spectra of the BNP41 glass-ceramics measured at different temperature. A series of semicircles are observed when the measuring temperature is above 420°C. As seen in Figure 8(b), the area of the semicircle decreases obviously as the measuring temperature increases, which means that the resistance of the grain boundary decreased. This decrease may be associated with the thermally activated motions of defect due to Maxwell–Wagner interfacial relaxation, which takes shape between the grains and the glass matrix.[7]

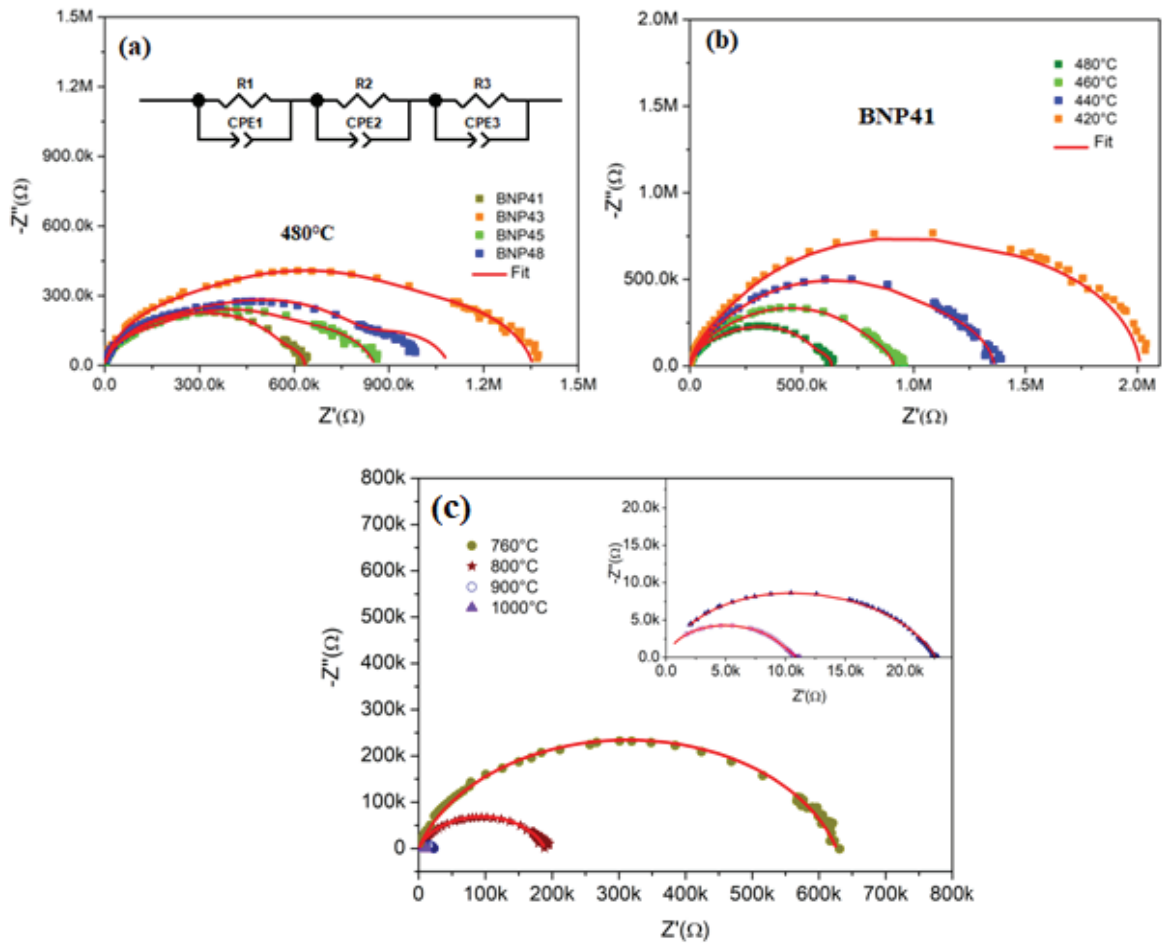

Figure 8. (a) Complex plane impedance plots ($Z'$–$Z''$) for BNPN-GC at 480°C, (b) Complex plane impedance plots ($Z'$–$Z''$) for BNP41 at various temperatures, (c) the impedance plots of the BNP41 crystallised at different temperature. The inset shows the equivalent electric circuit [Colour available online]

Table 2. *Resistive and capacitive parameters of model equivalent circuit for different samples estimated by the ZView software. Uncertainties of the fitted parameters are in brackets*

| Samples | $R_1$ (KΩ) | $C_1$ (pF) | $R_2$ (KΩ) | $C_2$ (pF) | $R_3$ (KΩ) | $C_3$ (pF) |
|---|---|---|---|---|---|---|
| BNP41 | 408·23 (213·5) | 135·8 (65·3) | 59·49 (20·6) | 4018 (1180) | 165·71 (122) | 152·3 (33) |
| BNP43 | 629·86 (185·6) | 74 (15) | 400·13 (110·5) | 1466 (687) | 326·37 (75·4) | 37·72 (8·6) |
| BNP45 | 416·3 (142) | 150·8 (46·1) | 279·75 (76·8) | 3463 (1845) | 159·86 (45·86) | 45·32 (10·11) |
| BNP48 | 557·35 (245·65) | 192·2 (87·5) | 218·82 (98·45) | 3814 (1856) | 283·3 (108·7) | 51·7 (10·5) |

The impedance plot of the BNP41 crystallised at different temperatures is shown in Figure 8(c). The corresponding equivalent circuit is modeled using two RC parallel circuits in series to simulate the effects of the grain and grain boundary. It can be clearly observed that the resistance decreases with increasing crystallisation temperature, which is related to the decrease of grain boundary resistance of glass-ceramics.[37] Moreover, a significant decrease in resistance was observed for 900 and 1000°C. Therefore, by increasing the crystallisation temperature, a higher conductivity could be obtained. Furthermore, this drop could be due to a reduction in the amount of non-amorphous phase compared to the crystalline phase. Indeed, the resistance of a residual glassy phase is often higher than that of a ceramic phase.[38]

In order to determine the activation energy, the relaxation time can be obtained from the relaxation frequency $f_r$ corresponding to the maximal value of peak of Z″, i.e. $2\pi f_r\tau=1$. According to the relation $\tau=\tau_0\exp(E_a/k_BT)$, the plot of versus $1/T$ presents a straight line as follows:

$$\ln \tau = \ln \tau_0 \exp\left(\frac{E_a}{k_B T}\right) \quad (1)$$

Where $\tau$, $\tau_0$, $E_a$, $k_B$ and $T$ denote the relaxation time, pre-exponential factor, activation energy, Boltzmann constant and the absolute temperature, respectively. The relaxation time $\tau$ is correlated to the impedance test frequency and is calculated by the following formula:

$$\tau = \frac{1}{\omega_{max}} = \frac{1}{2\pi f_{max}} \quad (2)$$

where $f_{max}$ is the frequency corresponding to the maximum impedance value. Figure 9 shows the frequency-dependence of imaginary part in impedance ($Z''$) at different measuring temperature for all the BNPN glass-ceramics. We can see that the extreme value of the $-Z''$ corresponding to relaxation frequency increases as the temperature increases, which indicates the temperature-dependent relaxation process in glass-ceramics.[39] It can be seen also that the $Z''$ maximum moves to a higher frequency as the measurement temperature increases, which indicates the reduction of relaxation time in the glass-ceramics.

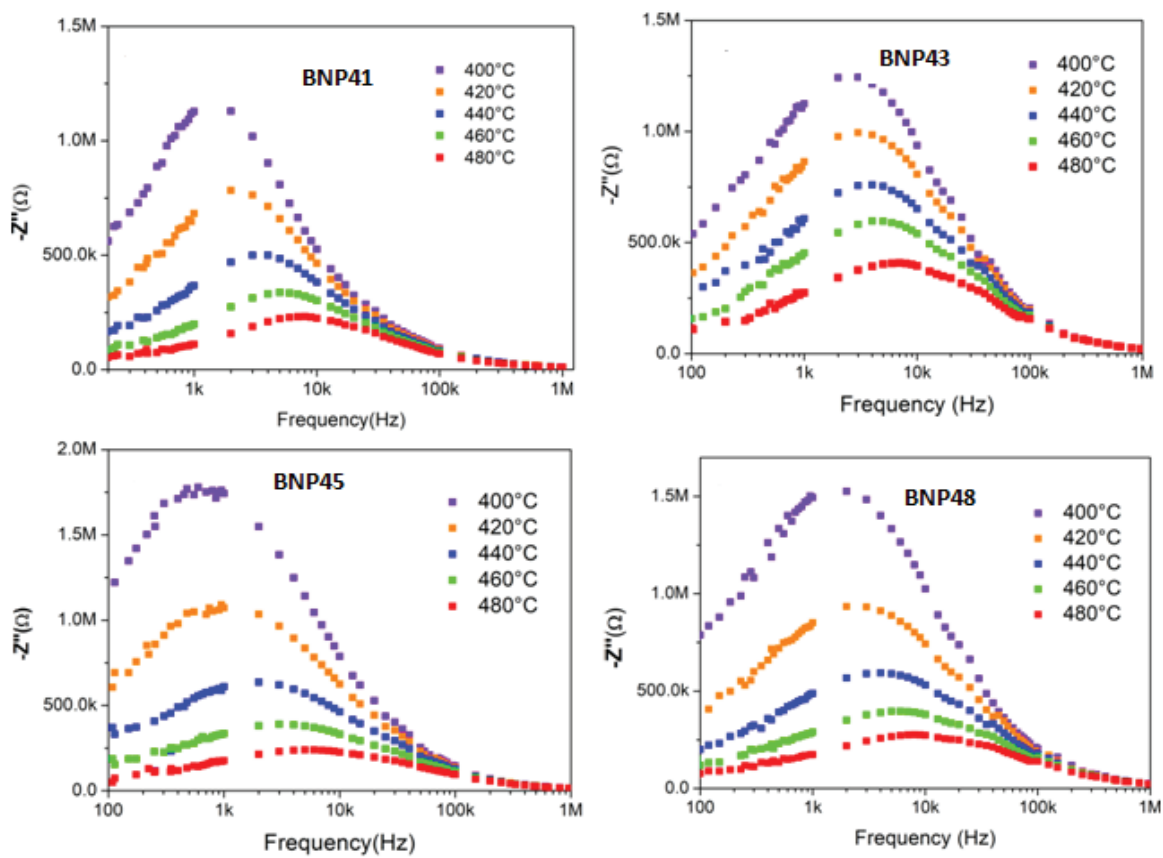

Figure 9. *Imaginary part of impedance (−Z″) as a function of frequency at different measuring temperature of glass-ceramics [Colour available online]*

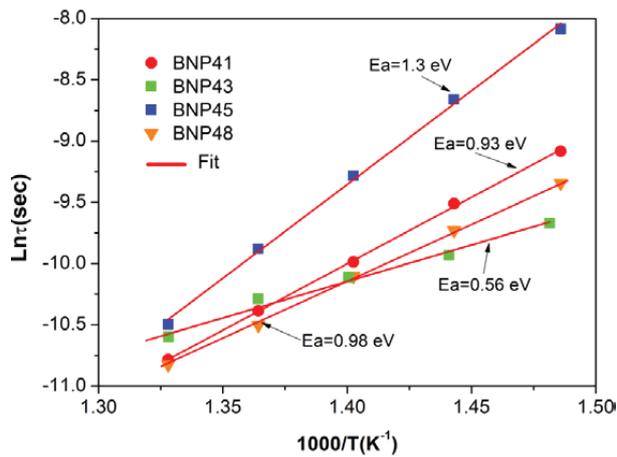

Figure 10. Relaxation time versus reciprocal of temperature (ln τ–1000/T) [Colour available online]

Table 3. Dielectric and energy storage properties of BNP41 crystallised at different temperatures

| T (°C) | $\varepsilon_r$ | tan δ | $E_a$ (eV) | $W_t$ (mJ/cm³) | $W_{rec}$ (mJ/cm³) | η (%) |
|---|---|---|---|---|---|---|
| 760 | 74 | 0·005 | 0·93 | 50·7 | 33·1 | 65·2 |
| 800 | 98 | 0·02 | 1·24 | 54·11 | 42·9 | 79·28 |
| 900 | 196 | 0·007 | 1·43 | 194·48 | 77·6 | 39·9 |
| 1000 | 276 | 0·011 | 1·57 | 205·1 | 105·8 | 51·6 |

The activation energy is calculated from Equation (1). The plots of ln τ–1000/T are shown in Figure 10. The calculated activation energy of BNPN glass-ceramics firstly drops down from 0·93 eV for x=0·41 to 0·56 eV for x=0·43, and then increases to 1·3 eV for x=0·45 and to 0·98 eV for x=0·48. The dielectric difference between the ferroelectric phase and the glass matrix causes charges accumulation at the interface. The value of $E_a$ represents the carrier propagating behavior through interface.[40] The lower activation energy is obtained for BNP43 sample, which could be explained by his means a weak interfacial po-

larisation at the interfaces. This could lead to a high breaking strength value (BDS).[7,40]

The $E_a$ values of the glass BNP41 crystallised at 760, 800, 900 and 1000°C were 0·93, 1·43, 1·24 and 1·57 eV, respectively. It is noted that the activation energy increase with the increase of crystallisation temperature. For BNPN41 crystallised at 1000°C, the activation energy could be related to the highly resistive crystalline phase and therefore a higher value (1·57 eV) is observed.

### 3.6. Energy storage properties

The polarisation–electric field (P–E) hysteresis loops of the BNNP ceramics at different crystallisation temperatures, measured at 1 kHz and 25°C are shown in Figure 11. The energy storage properties of the dielectric materials are characterised by discharge energy density ($W_{rec}$), energy loss density ($W_{loss}$), and discharge efficiency (η).[5]

The polarisation displays an increasing behaviour with the increase of crystallisation temperature under the same electric field. This result agrees with trend of variation of the dielectric constant. As shown in Figure 6, compared with all the P–E loops, no clear ferroelectric behaviour could be depicted. Therefore, the BNNP glass-ceramics were regarded as a linear dielectric material.

Dielectric and energy storage properties for $\varepsilon_r$, tan δ, $E_a$, $W_t$, $W_{rec}$ and η are listed in Table 3. As can be seen, the discharged energy density displays an increase tendency with increasing crystallisation temperature. For example, the discharged energy density of the glass-ceramics crystallised at 760°C increases from 33 to 105·8 mJ/cm³ for 1000°C under and electric field of 120 kV/cm. It is interesting to note that the discharge energy storage values obtained in this work are comparable to other silicate glass ceramics reported in the literature.[9]

Under the same electric field, the discharged energy efficiency (η) increases and then decreases with increasing crystallisation temperature. However, the energy efficiency increases with increasing crystallisation temperature, i.e. it increases from 65·2% for 760°C to 79·28% for 800°C under an applied electric field of 120 kV/cm and then decreases from 79·28 for 800°C to 39·9 and 51·6 for 900°C and 1000, respectively. This result can mainly be attributed to the increase of the interfacial polarisation at interfaces between the crystal phase and the glass matrix.[10] Noting that due to the increase of interfacial polarisation, more stored charges cannot be released in discharged process, which leads to the decrease of discharged energy efficiency.

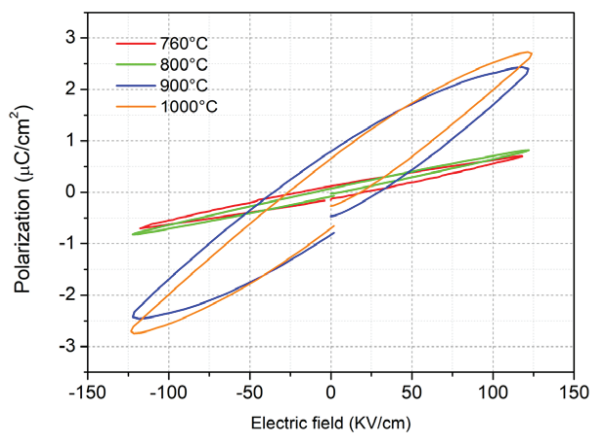

Figure 11. PE hysteresis loops for glass-ceramics studied at 120 kV/cm [Colour available online]

## 4. Conclusion

Lead-free (BaO, Na$_2$O)–Nb$_2$O$_5$–P$_2$O$_5$ were prepared successfully by controlling crystallisation method. The XRD analysis showed the existence of Ba$_2$NaNb$_5$O$_{15}$–NaNbO$_3$ co-crystallised phase in the samples. The content of Nb$_2$O$_5$ affected the dielectric properties of the glass-ceramics by changing their ferroelectric phase proportions. This study reveals that as the crystallisation temperature increases, the relative permittivity and energy storage density increases gradually because of the precipitation of the phases with high relative permittivity. The optimal dielectric constant of about 75 and 277 was obtained for BNP41 crystallised at 760 and 1000°C, respectively. The recoverable energy density of BNP41 crystallised at 1000°C reaches 105·8 mJ/cm$^3$ under an applied electric field of 120 kV/cm. It is founded that the energy efficiency depends on the interface polarisation with an optimum value of 78·7% obtained for 800°C under the electric field of 120 kV/cm at 1 kHz. The BNPN glass-ceramics possessing such high dielectric constant are deemed to open possible applications for the energy storage capacitor.

## Acknowledgments

The authors gratefully acknowledge the financial support of CNRST, OCP foundation and the European Union's Horizon H2020-MSCA-RISE research and innovation actions, ENGIMA and MELON.